# DETECTION AND CLASSIFICATION OF BREAST CANACER METASTATES BASED ON U-NET

*LIN XU, CHENG XU, YI TONG, YU CHUN SU*

GENERATIONSE SOFTWARE SOLUTIONS, INC., CANADA

## ABSTRACT

This paper presents U-net based breast cancer metastases detection and classification in lymph nodes, as well as patient-level classification based on metastases detection. The whole pipeline can be divided into five steps: preprocessing and data argumentation, patch-based segmentation, post processing, slide-level classification, and patient-level classification. In order to reduce overfitting and speedup convergence, we applied batch normalization and dropout into U-Net. The final Kappa score reaches 0.902 on training data.

## 1. INTRODUCTION

Deep learning has been successfully applied to many tasks including image processing [1], sound/voice processing [2], language translation [2]. Recent progress shows that deep learning can also be applied into medical image processing, such as MRI [3], CT, biopsy, endoscopy. In the field of pathology, most of work still need to be done manually by pathologist using microscopes. Therefore, such tasks are time-consuming and the results vary significantly depends on the skill and condition of individual pathologist. In particular, cancer metastases detection is very challenging and requires extensive microscopic assessment by pathologists as the size of tumor region can be very small.

An automated solution would potentially reduce the workload of pathologists as well as reduce the subjectivity in diagnosis [4]. The goal of Camelyon17 is automated detection and classification of breast cancer metastases in whole-slide images of histological lymph node. As the size of tumor regions can be very small (< 200 cells), pathologists are often required using high magnification (20X ~ 40X) for detecting tumor cells. This requirement significantly increases the workload for pathologist and machine learning approaches as well.

Although the goal of Camelyon17 Challenge is to determine a pN-stage for every patient in the test dataset based on sizes/numbers of positive lymph nodes in multiple slides (Macro: metastases > 2.0 mm; Micro: metastases > 0.2 mm / 200 cells, < 2.0 mm), the foundation of such tasks is to accurate segment cancer cells from normal cells. As the size of biopsy digital image is very large, there is no way to segment a whole image in a single pass, patch-based training/test pipeline becomes standard for handling Gigapixel pathology images. There are two common approaches for addressing this segmentation problem. The first approach solves metastases segmentation problem by classification [5] [6]. The whole slide is divided into multiple overlapping patches. Patch with more than half cancer cells is labelled as a cancer patch (positive), otherwise as a normal patch (negative). Therefore, metastases segmentation can be generated by averaging over classification results for all patches. The second approach is based on patch segmentation [7]. It still generates many patches, then segment cancer cells from normal ones for each patch. In this work, we choose the follow second approach as metastases can be very small and may not be the major component in some patches.

## 2. U-NET ARCHITECTURE

Our segmentation (pixel-wise classification) is based on U-Net architecture [8] as shown in Figure 1.

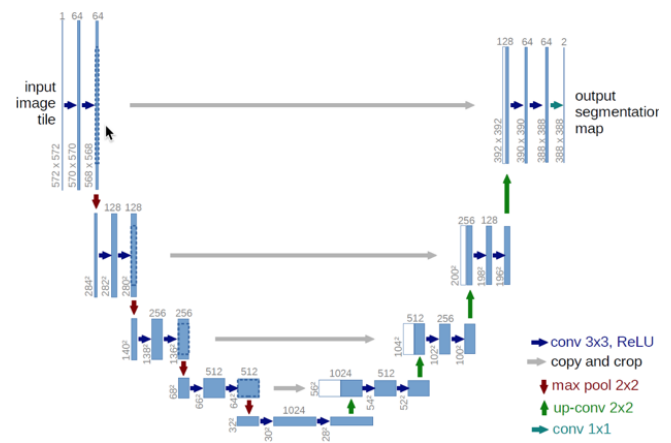

Figure 1: Original U-Net architecture

We made some necessary modifications in order to better fit our purpose. We changed the model input size from 572*572*1 to 256*256*3 in order to accept RGB images. Consequently, the following layers are also altered accordingly. Since Camelyon16 and Camelyon17 only provides less than 200 positive slides with masks, we also added batch normalization and dropout to the original implementation of U-Net in order to reduce overfitting and speedup convergence.

## 3. TRAINING PIPELINE

Our model training pipeline can be divided into 5 steps: preprocessing and data argumentation, patch-based segmentation, post processing, slide-level classification, and patient-level classification.

### 3.1. Preprocessing and data argumentation

As many deep learning-based approaches, many efforts have been made on data processing. We used all negative slides and positive slides from Camelyon16 and Camelyon17 with masks. Before taking samples from original slides, we generated regions of interest (ROI) by combining results from contours and Otsu's method [9] as shown in Figure 2.

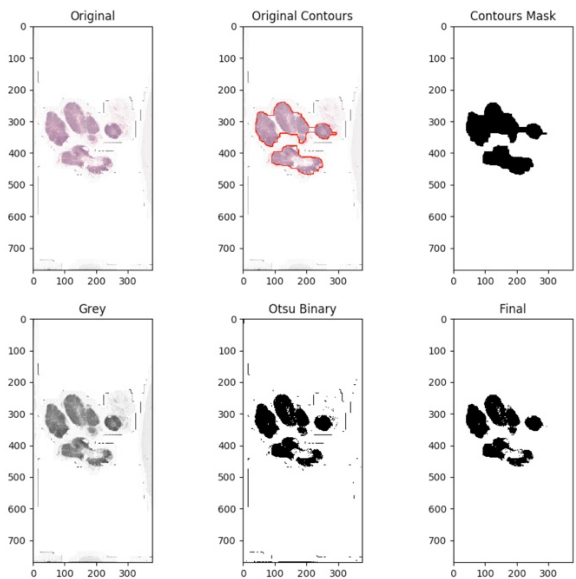

Figure 2: ROI detecting based on contours and Otsu

We selected *N* patches of size 512*512 that covered all tumor regions, then randomly selected *4*N* patches from normal regions from tumor slides as well as normal sides within region of interest (ROI). In the end, we oversampled tumor patches to *2*N*. Therefore, the negative/positive ratio was 2. In order to add variability into training data, heavy data argumentation was used such as flip, HSV color argumentation. Finally, all patches were resized to size 256*256.

### 3.2. Patch-based segmentation

We chose to use TensorFlow [10] implementation of U-Net. All positive and negative patches were sampled in advance; data argumentation was performed during training on the fly. Training data samples are shown in Figure 3.

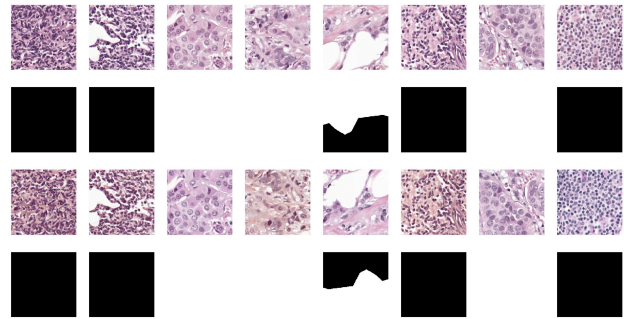

Figure 3: Training patch and corresponding mask without (top) and with argumentation (bottom)

We trained our model on Dell T630 server with 128G memory and 4 Titan X GPUs. Initially, the learning rate was set to 0.001, and reduced to 0.0001 after 20 epochs. The final segmentation performance is evaluated on independent validation slides (see Figure 4).

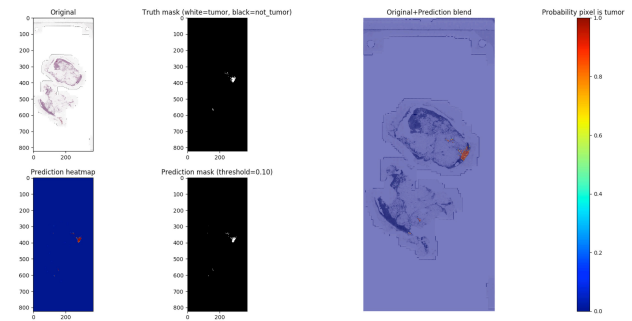

Figure 4: Sample segmentation results

### 3.3. Post processing

The segmentation produces a heat mask with probability for each pixel being cancer/normal for each slide. We empirically selected a threshold value to separate tumor cells from normal ones. We also conducted experiments with average filter to filter out noise. However, applying such filter may remove small isolated tumor regions. Therefore, we did not use average filter for slide-level classification.

## 3.4. Slide-level classification

After post processing, we obtained tumor regions. We adopted random forest approach with 16 hand crafted features mainly focused on the size of the tumor region. In order to reduce the bias introduced by manually selecting heat map threshold, our features are based on three different thresholds, namely 0.3, 0.5, 0.7, 0.9.

Figure 5 shows confusion matrix on validation data. It is worth to note that classification accuracy for ITC (tumor regions with size < 0.2 mm or < 200 cells) is very low, and ITC almost always be misclassified as negative.

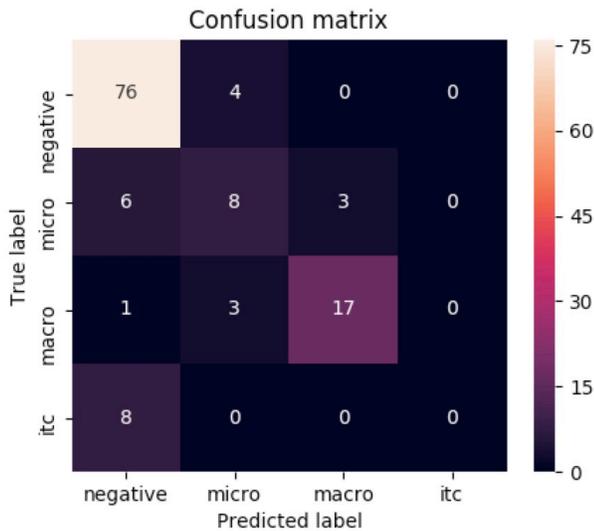

Figure 5: confusion matrix for slide-level classification on validation data (126 slides)

## 3.5. Patient-level classification

We did not perform additional training for patient-level classification. Rather, we followed simple Pathologic lymph node classification (pN-stage) standard defined in CAMELYON17 [4].

- pN0: No micro-metastases or macro-metastases or ITCs found.
- pN0(i+): Only ITCs found.
- pN1mi: Micro-metastases found, but no macro-metastases found.
- pN1: Metastases found in 1–3 lymph nodes, of which at least one is a macro-metastasis.
- pN2: Metastases found in 4–9 lymph nodes, of which at least one is a macro-metastasis.

## 4. SUMMARY OF UPDATE

After initial submission, we revisited the segmentation results on validation data. The observations suggested following updates on this version.

- We retrained our mode with updated data from Camelyon.
- Instead of working on 40X magnification, we chose 20X magnification for extracting patches for training and test.
- We removed rotation-based augmentation and only applied HSV, brightness, flip based augmentations.
- In order to speedup training, we selected all tumor slides with label (mask) from Camelyon17 and Camelyon16, and only selected normal slides from Cameylon17. Overall, we chose 209 tumor sides and 313 normal sides.
- We also applied oversampling on positive patches but also kept positive/negative ration as 3. Therefore, we introduced more negative samples to reduce false positive.
- We included Camelyon16 testing slides as training data for slide-level classification.

For slide-level classification, we introduced more features based on negative-density of biggest predict tumor region to capture neighborhood information.

## 5. RESULT AND CONCLUSION

We tested our pipeline on validation data and obtained Kappa score of 0.902. The final Kappa score is computed using scripts provide by Camelyon17.

Obviously, there are many room for improvement. We noticed that ITC is very hard to predict. This may due to the patterns of isolated tumor cells are somewhat unique, and there are not enough training data for the model to capture such pattern. By random sampling, we did not obtain enough negative samples on such special regions. Therefore, we should perform important sampling or hard mining to introduce more, harder patches into training data and make the model more focus on mistakes. Another way to improve model prediction accuracy can be applying model assembly. We are working on training models with different architectures, even combining classification approach with segmentation.

Overall, computer-aided diagnosis system will be one of the most useful applications of deep learning. A technology that improves the detection of tumors for breast cancer can be easily adopted for detecting many other types of diseases.